\documentclass[12pt,a4paper]{article}

\usepackage{afterpage}
\usepackage{color}
\usepackage[normalem]{ulem}

\usepackage{ifthen} 
\newboolean{pdflatex}
\setboolean{pdflatex}{true} 

\newboolean{articletitles}
\setboolean{articletitles}{true} 

\newboolean{uprightparticles}
\setboolean{uprightparticles}{true} 

\newboolean{inbibliography}
\setboolean{inbibliography}{false}

\usepackage{microtype}
\usepackage{lineno}  
\usepackage{xspace} 

\usepackage{comment}

\usepackage{graphicx}  
\usepackage{color}
\usepackage{colortbl}
\graphicspath{{./figs/}} 

\usepackage{graphpap}
\usepackage{rotating}

\usepackage{multirow}
\usepackage{verbatim}

\usepackage{amsmath} 
\usepackage{amssymb}
\usepackage{amsfonts}
\usepackage{upgreek} 

\newcommand*\patchAmsMathEnvironmentForLineno[1]{%
\expandafter\let\csname old#1\expandafter\endcsname\csname #1\endcsname
\expandafter\let\csname oldend#1\expandafter\endcsname\csname
end#1\endcsname
 \renewenvironment{#1}%
   {\linenomath\csname old#1\endcsname}%
   {\csname oldend#1\endcsname\endlinenomath}%
}
\newcommand*\patchBothAmsMathEnvironmentsForLineno[1]{%
  \patchAmsMathEnvironmentForLineno{#1}%
  \patchAmsMathEnvironmentForLineno{#1*}%
}
\AtBeginDocument{%
\patchBothAmsMathEnvironmentsForLineno{equation}%
\patchBothAmsMathEnvironmentsForLineno{align}%
\patchBothAmsMathEnvironmentsForLineno{flalign}%
\patchBothAmsMathEnvironmentsForLineno{alignat}%
\patchBothAmsMathEnvironmentsForLineno{gather}%
\patchBothAmsMathEnvironmentsForLineno{multline}%
}

\usepackage{hyperref}    
\usepackage[all]{hypcap} 

\def\lhcb {\mbox{LHCb}\xspace}

\def\babar  {\mbox{BaBar}\xspace}
\def\belle  {\mbox{Belle}\xspace}

\def\cdf    {\mbox{CDF}\xspace}

\def\spd    {SPD\xspace}
\def\presh  {PS\xspace}
\def\ecal   {ECAL\xspace}

\ifthenelse{\boolean{uprightparticles}}%
{
 
 \def\Pgamma      {\ensuremath{\upgamma}\xspace}

 \def\Pvarepsilon {\ensuremath{\upvarepsilon}\xspace}                 
                  
 \def\Peta        {\ensuremath{\upeta}\xspace}

 \def\Pmu         {\ensuremath{\upmu}\xspace}

 \def\Ppi         {\ensuremath{\uppi}\xspace}

 \def\Pchi        {\ensuremath{\upchi}\xspace}                 
 \def\Ppsi        {\ensuremath{\uppsi}\xspace}

 \def\PDelta      {\ensuremath{\Delta}\xspace}                 
 \def\PXi      {\ensuremath{\Xi}\xspace}                 
 \def\PLambda      {\ensuremath{\Lambda}\xspace}                 
 \def\PSigma      {\ensuremath{\Sigma}\xspace}                 
 \def\POmega      {\ensuremath{\Omega}\xspace}                 
 \def\PUpsilon      {\ensuremath{\Upsilon}\xspace}                 
 
 \def\PB      {\ensuremath{\mathrm{B}}\xspace}                 
                  
 \def\PD      {\ensuremath{\mathrm{D}}\xspace}

 \def\PJ      {\ensuremath{\mathrm{J}}\xspace}                 
 \def\PK      {\ensuremath{\mathrm{K}}\xspace}

 \def\PX      {\ensuremath{\mathrm{X}}\xspace}                 
 \def\PY      {\ensuremath{\mathrm{Y}}\xspace}

 \def\Pb      {\ensuremath{\mathrm{b}}\xspace}                 
 \def\Pc      {\ensuremath{\mathrm{c}}\xspace}                 
                  
 \def\Pe      {\ensuremath{\mathrm{e}}\xspace}                 
                  
 \def\Pg      {\ensuremath{\mathrm{g}}\xspace}                 
                  
 \def\Pi      {\ensuremath{\mathrm{i}}\xspace}

 \def\Pp      {\ensuremath{\mathrm{p}}\xspace}

}
{
 
 \def\Pgamma      {\ensuremath{\gamma}\xspace}

 \def\Pvarepsilon {\ensuremath{\varepsilon}\xspace}                 
                  
 \def\Peta        {\ensuremath{\eta}\xspace}

 \def\Pmu         {\ensuremath{\mu}\xspace}

 \def\Ppi         {\ensuremath{\pi}\xspace}

 \def\Pchi        {\ensuremath{\chi}\xspace}                 
 \def\Ppsi        {\ensuremath{\psi}\xspace}                 
                  
 \mathchardef\PDelta="7101
 \mathchardef\PXi="7104
 \mathchardef\PLambda="7103
 \mathchardef\PSigma="7106
 \mathchardef\POmega="710A
 \mathchardef\PUpsilon="7107
                  
 \def\PB      {\ensuremath{B}\xspace}                 
                  
 \def\PD      {\ensuremath{D}\xspace}

 \def\PJ      {\ensuremath{J}\xspace}                 
 \def\PK      {\ensuremath{K}\xspace}

 \def\PX      {\ensuremath{X}\xspace}                 
 \def\PY      {\ensuremath{Y}\xspace}

 \def\Pb      {\ensuremath{b}\xspace}                 
 \def\Pc      {\ensuremath{c}\xspace}                 
                  
 \def\Pe      {\ensuremath{e}\xspace}                 
                  
 \def\Pg      {\ensuremath{g}\xspace}                 
                  
 \def\Pi      {\ensuremath{i}\xspace}

 \def\Pp      {\ensuremath{p}\xspace}

}

\def\epem       {{\ensuremath{\Pe^+\Pe^-}}\xspace}

\def\mumu       {{\ensuremath{\Pmu^+\Pmu^-}}\xspace}

\def\g      {{\ensuremath{\Pgamma}}\xspace}

\def\cquark    {{\ensuremath{\Pc}}\xspace}
\def\cquarkbar {{\ensuremath{\overline \cquark}}\xspace}
\def\ccbar     {{\ensuremath{\cquark\cquarkbar}}\xspace}
\def\bquark    {{\ensuremath{\Pb}}\xspace}

\def\pion   {{\ensuremath{\Ppi}}\xspace}
\def\piz    {{\ensuremath{\pion^0}}\xspace}

\def\pip    {{\ensuremath{\pion^+}}\xspace}
\def\pim    {{\ensuremath{\pion^-}}\xspace}

\def\kaon    {{\ensuremath{\PK}}\xspace}
  \def\Kbar    {{\kern 0.2em\overline{\kern -0.2em \PK}{}}\xspace}

\def\Kp      {{\ensuremath{\kaon^+}}\xspace}

\def\Kstarp  {{\ensuremath{\kaon^{*+}}}\xspace}

\def\Dbar    {{\bar\PD}\xspace}
\def\D       {{\ensuremath{\PD}}\xspace}

\def\Dstarb  {{\ensuremath{\Dbar^*}}\xspace}

\def\B       {{\ensuremath{\PB}}\xspace}
\def\Bbar    {{\ensuremath{\kern 0.18em\overline{\kern -0.18em \PB}{}}}\xspace}

\def\Bu      {{\ensuremath{\B^+}}\xspace}

\def\Bp      {{\ensuremath{\Bu}}\xspace}

\def\jpsi     {{\ensuremath{{\PJ\mskip -3mu/\mskip -2mu\Ppsi\mskip 2mu}}}\xspace}
\def\psitwos  {{\ensuremath{\Ppsi{(2S)}}}\xspace}

\def\chicone  {{\ensuremath{\Pchi_{\cquark 1}}}\xspace}

  \def\Y#1S{\ensuremath{\PUpsilon{(#1S)}}\xspace}

\def\proton      {{\ensuremath{\Pp}}\xspace}

\def\Lbar        {{\ensuremath{\kern 0.1em\overline{\kern -0.1em\PLambda}}}\xspace}

\def\BF         {{\ensuremath{\cal B}}\xspace}

\def\BR         {\BF}

\def\to                 {\ensuremath{\rightarrow}\xspace}

\def\AT#1     {\ensuremath{A_{\mathrm{T}}^{#1}}\xspace}           

\def\C#1      {\ensuremath{\mathcal{C}_{#1}}\xspace}                       
\def\Cp#1     {\ensuremath{\mathcal{C}_{#1}^{'}}\xspace}                    
\def\Ceff#1   {\ensuremath{\mathcal{C}_{#1}^{\mathrm{(eff)}}}\xspace}        
\def\Cpeff#1  {\ensuremath{\mathcal{C}_{#1}^{'\mathrm{(eff)}}}\xspace}       
\def\Ope#1    {\ensuremath{\mathcal{O}_{#1}}\xspace}                       
\def\Opep#1   {\ensuremath{\mathcal{O}_{#1}^{'}}\xspace}                    

\newcommand{\tev}{\ifthenelse{\boolean{inbibliography}}{\ensuremath{~T\kern -0.05em eV}\xspace}{\ensuremath{\mathrm{\,Te\kern -0.1em V}}}\xspace}
\newcommand{\gev}{\ensuremath{\mathrm{\,Ge\kern -0.1em V}}\xspace}
\newcommand{\mev}{\ensuremath{\mathrm{\,Me\kern -0.1em V}}\xspace}
\newcommand{\kev}{\ensuremath{\mathrm{\,ke\kern -0.1em V}}\xspace}
\newcommand{\ev}{\ensuremath{\mathrm{\,e\kern -0.1em V}}\xspace}
\newcommand{\gevc}{\ensuremath{{\mathrm{\,Ge\kern -0.1em V\!/}c}}\xspace}
\newcommand{\mevc}{\ensuremath{{\mathrm{\,Me\kern -0.1em V\!/}c}}\xspace}
\newcommand{\gevcc}{\ensuremath{{\mathrm{\,Ge\kern -0.1em V\!/}c^2}}\xspace}
\newcommand{\gevgevcccc}{\ensuremath{{\mathrm{\,Ge\kern -0.1em V^2\!/}c^4}}\xspace}
\newcommand{\mevcc}{\ensuremath{{\mathrm{\,Me\kern -0.1em V\!/}c^2}}\xspace}

\def\mum  {\ensuremath{{\,\upmu\rm m}}\xspace}

\def\invfb   {\ensuremath{\mbox{\,fb}^{-1}}\xspace}

\newcommand{\chisq}{\ensuremath{\chi^2}\xspace}

\def\gsim{{~\raise.15em\hbox{$>$}\kern-.85em
          \lower.35em\hbox{$\sim$}~}\xspace}
\def\lsim{{~\raise.15em\hbox{$<$}\kern-.85em
          \lower.35em\hbox{$\sim$}~}\xspace}

\def\pt         {\mbox{$p_{\rm T}$}\xspace}

\def\evtgen     {\mbox{\textsc{EvtGen}}\xspace}

\def\geant      {\mbox{\textsc{Geant4}}\xspace}

\def\photos     {\mbox{\textsc{Photos}}\xspace}

\def\pythia     {\mbox{\textsc{Pythia}}\xspace}

\def\tell1  {TELL1\xspace}
\def\ukl1   {UKL1\xspace}

\def\X{\ensuremath{\PX(3872)}\xspace}
\def\psipr{\ensuremath{\Ppsi(2\PS)}\xspace}

\usepackage{cite} 
\usepackage{mciteplus}

\usepackage{longtable} 

\renewcommand{\psitwos}{\ensuremath{\Ppsi{\mathrm{(2S)}}}\xspace}
\renewcommand{\psipr}  {\ensuremath{\Ppsi{\mathrm{(2S)}}}\xspace}

\begin{document}

\renewcommand{\thefootnote}{\fnsymbol{footnote}}
\setcounter{footnote}{1}


\begin{titlepage}
\pagenumbering{roman}

\vspace*{-1.5cm}
\centerline{\large EUROPEAN ORGANIZATION FOR NUCLEAR RESEARCH (CERN)}
\vspace*{1.5cm}
\hspace*{-0.5cm}
\begin{tabular*}{\linewidth}{lc@{\extracolsep{\fill}}r}
\ifthenelse{\boolean{pdflatex}}
{\vspace*{-2.7cm}\mbox{\!\!\!\includegraphics[width=.14\textwidth]{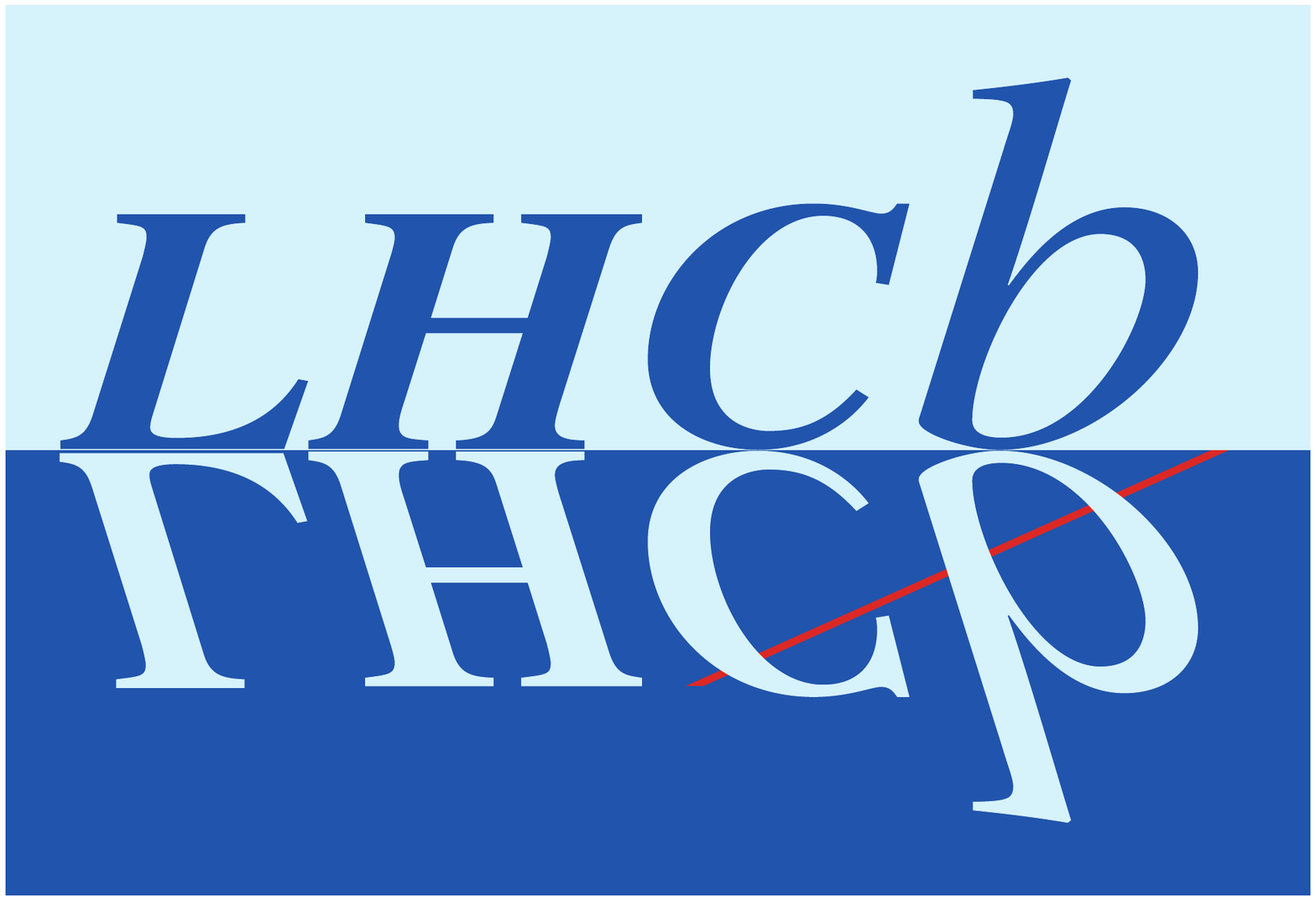}} & &}%
{\vspace*{-1.2cm}\mbox{\!\!\!\includegraphics[width=.12\textwidth]{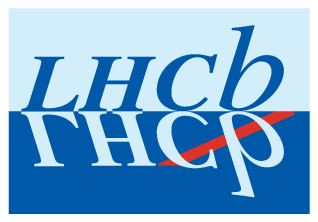}} & &}%
\\
 & & CERN-PH-EP-2014-050 \\  
 & & LHCb-PAPER-2014-008 \\  
 & & April, 1, 2014 \\ 
\end{tabular*}

\vspace*{2.0cm}

{\bf\boldmath\huge
\begin{center}
 Evidence for the decay $\X\to\psipr\g$
\end{center}
}

\vspace*{0.5cm}

\begin{center}
The LHCb collaboration\footnote{Authors are listed on the following pages.}
\end{center}

\vspace{0.5cm}

\begin{abstract}
Evidence for the decay mode $\X\to\psipr\g$ in $\Bp\to\X\Kp$ decays is found with a significance of 4.4 standard deviations. 
The analysis is based on a~data sample of proton-proton collisions,
corresponding to an integrated luminosity of~3\invfb,
collected with the \lhcb detector,
at centre-of-mass energies of 7~and~8\tev. 
The ratio of the branching fraction of the $\X\to\psipr\g$ decay to that of
the~$\X\to\jpsi\g$ decay is measured to be
\begin{equation*}
\dfrac{\BR(\X\to\psipr\g)}{\BR(\X\to\jpsi\g)} = 2.46\pm0.64\pm0.29, 
\end{equation*}
where the first uncertainty is statistical and the second is systematic.
The measured value agrees with expectations for a pure charmonium interpretation
of the~\X~state and a~mixture of charmonium and molecular interpretations. 
However, it does not support a pure~$\D\Dstarb$~molecular interpretation of the \X~state. 
\end{abstract}

\vspace*{2.0cm}

\begin{center}
Submitted to Nucl.~Phys.~B
\end{center}

\vspace{\fill}

{\footnotesize 
\centerline{\copyright~CERN on behalf of the \lhcb collaboration, license \href{http://creativecommons.org/licenses/by/3.0/}{CC-BY-3.0}.}}
\vspace*{2mm}

\end{titlepage}


\newpage
\setcounter{page}{2}
\mbox{~}
\newpage

\centerline{\large\bf LHCb collaboration}
\begin{flushleft}
\small
R.~Aaij$^{41}$, 
B.~Adeva$^{37}$, 
M.~Adinolfi$^{46}$, 
A.~Affolder$^{52}$, 
Z.~Ajaltouni$^{5}$, 
J.~Albrecht$^{9}$, 
F.~Alessio$^{38}$, 
M.~Alexander$^{51}$, 
S.~Ali$^{41}$, 
G.~Alkhazov$^{30}$
% ...

{\footnotesize \it
% ...
$ ^{5}$Clermont Universit\'{e}, Universit\'{e} Blaise Pascal, CNRS/IN2P3, LPC, Clermont-Ferrand, France\\
% ...
$ ^{9}$Fakult\"{a}t Physik, Technische Universit\"{a}t Dortmund, Dortmund, Germany\\
% ...
$ ^{30}$Petersburg Nuclear Physics Institute (PNPI), Gatchina, Russia\\
% ...
$ ^{37}$Universidad de Santiago de Compostela, Santiago de Compostela, Spain\\
$ ^{38}$European Organization for Nuclear Research (CERN), Geneva, Switzerland\\
% ...
$ ^{41}$Nikhef National Institute for Subatomic Physics, Amsterdam, The Netherlands\\
% ...
$ ^{46}$H.H. Wills Physics Laboratory, University of Bristol, Bristol, United Kingdom\\
% ...
$ ^{51}$School of Physics and Astronomy, University of Glasgow, Glasgow, United Kingdom\\
$ ^{52}$Oliver Lodge Laboratory, University of Liverpool, Liverpool, United Kingdom\\
% ...
}
\end{flushleft}

\cleardoublepage


\renewcommand{\thefootnote}{\arabic{footnote}}
\setcounter{footnote}{0}



\pagestyle{plain} 
\setcounter{page}{1}
\pagenumbering{arabic}


%

\section{Introduction}
\label{sec:Introduction}

The \X state was discovered in 2003 by the \belle collaboration~\cite{belleX}. 
Subsequently, it~has been studied by several other 
experiments~\cite{cdfX,dzeroX,babarX,lhcbX,cmsX_pipi}. 
Several properties of the \X state have been determined, including the~precise value of its mass~\cite{cdfX_mass,lhcbX} and
the~dipion mass spectrum 
in the~decay $\X\to\jpsi\pip\pim$~\cite{belleX,cdfX_pipi,cmsX_pipi}. 
Recently, its quantum numbers were determined to be $J^{PC}=1^{++}$ by combination
of the~measurements performed by the~\cdf~\cite{cdf_X3872_numb} and the~\lhcb~\cite{LHCb-PAPER-2013-001}~collaborations.

Despite a~large amount of experimental information, the nature of \X~state
and other similar states is still uncertain~\cite{XYZ,XYZ_proc}.
In particular for the~\X state, interpretation as
a~$\D\Dstarb$~molecule~\cite{moleculaX},
tetraquark~\cite{tetraX},
$\ccbar\Pg$ hybrid meson~\cite{hybridX},
vector glueball~\cite{glueballX} or mixed \mbox{state~\cite{Xmixture2,Xmixture1}} are proposed.
Radiative decays of the $\X$ provide a~valuable opportunity to understand its nature. 
Studies of the decay modes \mbox{$\X\to\jpsi\g$} resulted in the determination of its ${C\text{-parity}}$~\cite{babar_jpsi,belle}. 
Evidence for the \mbox{$\X\to\psipr\g$} decay and the branching fraction ratio,
\begin{equation*}
R_{\Ppsi\g} \equiv \frac{\BR(\PX(3872)\to\psitwos\g)}{\BR(\PX(3872)\to\jpsi\g)} = 3.4 \pm 1.4,
\end{equation*}
were reported by the \babar collaboration~\cite{babar}. 
In contrast, no significant signal was found for the ${\X\to\psipr\g}$ decay by the \belle collaboration, therefore 
only an upper limit for ${R_{\Ppsi\g} < 2.1~(\text{at 90\% confidence level})}$ was reported~\cite{belle}.
The ratio $R_{\Ppsi\g}$ is predicted to be in the range $( 3 - 4 )\times10^{-3}$ for a~$\D\Dstarb$~molecule~\cite{Swanson,DongFaeser,FerrettiGalata}, 
$1.2 - 15$ for a pure charmonium state~\cite{BarnesGodfreySwanson,BarnesGodfrey,LiChao,Lahde,Simonov,identifyX1,identifyX2} 
and $0.5 - 5$ for a~molecule-charmonium mixture~\cite{Simonov,EichtenLaneQuigg}.

In this paper, evidence for the decay $\X\to\psipr\g$ and a measurement of the~ratio
$R_{\Ppsi\g}$ using $\Bp\to\X\Kp$ decays are presented.\footnote{The inclusion of charged conjugate processes is implied throughout.} 
The analysis is based on a~data sample of proton-proton ($\proton\proton$)~collisions,
corresponding to an integrated luminosity of  1\invfb at a~centre-of-mass energy of $7\tev$
and 2\invfb at $8\tev$, collected with the~\lhcb detector.

\section{Detector and software}
\label{sec:Detector}

The \lhcb detector~\cite{Alves:2008zz} is a single-arm forward
spectrometer covering the \mbox{pseudorapidity} range $2<\Peta <5$,
designed for the study of particles containing $\mathrm{b}$ or $\mathrm{c}$
quarks. The~detector includes a high-precision tracking system
consisting of a silicon-strip vertex detector surrounding the $\mathrm{pp}$
interaction region, a large-area silicon-strip detector located
upstream of a dipole magnet with a bending power of about
$4{\rm\,Tm}$, and three stations of silicon-strip detectors and straw
drift tubes placed downstream.
The combined tracking system provides a momentum measurement with
relative uncertainty that varies from 0.4\,\% at 5\gevc to 0.6\,\% at 100\gevc,
and impact parameter resolution of 20\mum for
tracks with high transverse momentum. Charged hadrons are identified
using two ring-imaging Cherenkov detectors~\cite{LHCb-DP-2012-003}. 
The calorimeter system consists of a scintillating
pad detector (\spd) and a~pre-shower system (\presh), followed by
electromagnetic (\ecal) and hadron calorimeters. 
The \spd and \presh are designed to distinguish between signals from photons and electrons.
Muons are identified by a system composed of alternating layers 
of iron and multiwire proportional chambers~\cite{LHCb-DP-2012-002}.
 


The trigger~\cite{LHCb-DP-2012-004} consists of a hardware stage, based on information from the calorimeter
and muon systems, followed by a software stage where a full event reconstruction is applied. 
Events are first required to pass the hardware trigger, which selects muons with a transverse momentum, \pt, greater than 1.48\gevc. 
In the subsequent software trigger, at least one of the final state particles is required to have both $\pt>0.8\gevc$ and impact parameter in excess of $100\mum$ 
with respect to all of the primary $\proton\proton$ interaction vertexes~(PVs) in the~event.
Finally, the tracks of two or more of the final state 
particles are required to form a~vertex that is significantly displaced from the~PVs.

The analysis technique reported below has been validated using simulated events. The $\proton\proton$ collisions are generated using 
\pythia~\cite{Sjostrand:2006za,*Sjostrand:2007gs}  with a specific \lhcb configuration described in Ref.~\cite{LHCb-PROC-2010-056}. 
Decays of hadronic particles are described by \evtgen~\cite{Lange:2001uf} in which final state radiation
is generated using \photos~package~\cite{Golonka:2005pn}. The~interaction of the~generated particles with
the~detector and its response are implemented using the~\geant toolkit~\cite{Agostinelli:2002hh,Allison:2006ve} as described
in Ref.~\cite{LHCb-PROC-2011-006}.

\section{Event selection}
\label{sec:EventSelection}

Candidate $\Bp\to\X\Kp$~decays, followed by $\X\to\Ppsi\g$, where $\Ppsi$ 
denotes a \jpsi or \psipr meson, are reconstructed using the $\Ppsi\to\mumu$ channel. 
The $\psipr\to\jpsi\pip\pim$ decay mode is not used due to low reconstruction efficiency.
Most selection criteria are common for the two channels, except where directly related to the photon kinematics,
due to the difference in the energy release in these two channels. 
The selection criteria follow those used in 
Refs.~\cite{LHCb-PAPER-2013-024,LHCb-PAPER-2012-022,LHCb-PAPER-2012-053}. 

The track quality of reconstructed charged particles is ensured by requiring that the~$\chisq$ per degree of freedom, $\chi^2/\mathrm{ndf}$, is less than~3. 
Well-identified muons are selected by requiring that the difference in the logarithms of the muon hypothesis likelihood with respect
to the pion hypothesis  likelihood, $\Delta\log \mathcal{L}_{\mu/\pion}$~\cite{LHCb-DP-2013-001}, is larger than zero. 
To select kaons, the corresponding difference in the logarithms of likelihoods of
the kaon and pion hypotheses~\cite{LHCb-DP-2012-003} is required to satisfy $\Delta\log \mathcal{L}_{\kaon/\pion} > 0$.

To ensure that the muons and kaons do not originate from a $\mathrm{pp}$ interaction vertex,
the~impact parameter $\chisq$, defined as 
the difference between the $\chisq$ of a given PV formed with and without the considered track, is required to be $\chisq_{\mathrm{IP}} > 4$.
When more than one PV is reconstructed, the smallest value of $\chisq_{\mathrm{IP}}$ is chosen. 

Pairs of oppositely charged tracks identified as muons, each having $\pt>0.55\gevc$, are combined to form $\Ppsi\to\mumu$ candidates. 
The fit of the common two-prong vertex is required to satisfy $\chisq<20$. The vertex is required to be well separated from
the~reconstructed PV by selecting candidates with decay length significance greater than~3. 
The~invariant mass of the dimuon combination is required to be between 3.020~and~3.135\gevcc for
the~\jpsi candidates and between 3.597~and~3.730\gevcc for the \psipr candidates.

Photons are reconstructed using the~electromagnetic calorimeter and identified using a~likelihood-based estimator,
constructed from variables that rely on calorimeter and tracking information~\cite{LHCb-PAPER-2011-030}.  
Candidate photon clusters must not be matched to the~trajectory of a~track extrapolated from
the~tracking system to the~cluster position in the~calorimeter.
Further photon quality refinement is done using information from the~\presh and \spd detectors. 
The photon transverse momentum is required to be greater than 1\gevc~or~0.6\gevc
for the~\jpsi or \psipr in the~final state, respectively. 
To suppress the large combinatorial background from ${\piz\to\g\g}$ decays, a pion veto is applied~\cite{LHCb-PAPER-2012-022}. 
The photons that, when combined with another photon, form a ${\piz\to\g\g}$ candidate with invariant mass
within $25\mevcc$ of the~$\piz$ mass,
corresponding to $\pm3$ times the~mass resolution~\cite{LHCb-PAPER-2012-022,Dasha_disser}, 
are not used in the~reconstruction. 

To form $\X$ candidates, the selected \Ppsi candidates are combined with a reconstructed photon.
To be considered as a $\X$ candidate, the $\jpsi\g$ or $\psipr\g$ combination must have an invariant
mass in the~range ${3.7\text{ -- }4.1}$\gevcc or ${3.75\text{ -- }4.05}$\gevcc, respectively, to account
for the~different available phase space.

The $\X$ candidates are combined with selected kaons to create \Bp meson candidates. The kaons are required to have transverse 
momentum larger than 0.8\gevc. The~quality of the \Bp~vertex is ensured by requiring the $\chisq$ of the 
vertex fit to be less than 25. In addition, the decay time of the \Bp is required to be larger than~150\mum/$c$ to reduce the large combinatorial 
background from particles produced at the~PV. 

To improve the invariant mass resolution of the \X candidate, a~kinematic fit~\cite{DTF} is performed. 
In this fit, the invariant mass of the $\Ppsi$ candidate is constrained to its nominal value~\cite{PDG2012}, 
the decay products of the \Bp candidate are 
required to originate from a~common vertex, and the momentum vector of the \Bp candidate is required to point back to the~PV.
The \chisq/ndf for this fit is required to be less than~5. 
To improve the resolution on  the~\Bp candidate invariant mass,
and minimize its correlation with the~reconstructed \X~candidate mass,
the~\Bp mass is determined from a similar kinematic fit with an~additional constraint applied 
to the mass of the~\X~resonance~\cite{PDG2012}.
The~\Bp~candidates are required to have invariant mass in the range ${5.0 - 5.5\gevcc}$. 
To~reject possible contributions from $\Bp\to\Ppsi\Kp$ decays with 
an additional random soft photon, the~invariant mass of the $\Ppsi\Kp$ combination is required 
to be outside a~$\pm40\mevcc$~mass window around the~known \Bp~mass~\cite{PDG2012}.


\section[ Signal yields ] { Signal yields }
\label{sec:signal}


To determine the signal yield of the ${\Bp\to\X\Kp}$ decays followed by ${\X\to\Ppsi\g}$, an unbinned extended maximum likelihood 
two-dimensional fit in ${\Ppsi\g\Kp}$ and $\Ppsi\g$ invariant masses is performed. 
The probability density function used in the fit consists of three components to 
describe the mass spectrum:~signal, background from other \B decays that peaks in the ${\Ppsi\g\Kp}$ and $\Ppsi\g$ invariant mass
distributions (henceforth called ``peaking background'') and combinatorial background. 
The signal component is modelled as a~product of a~Gaussian function 
in the ${\Ppsi\g\Kp}$ invariant mass and a Crystal~Ball function~\cite{CrystalBall1} in the  
$\Ppsi\g$ invariant mass. The mass resolution and tail parameters 
of the Crystal~Ball function are fixed to those determined from simulated signal events.

The peaking background is studied using simulation.
The sources of the peaking background are different in 
the \jpsi and \psipr channels due to differences in the photon spectra 
and in the photon selection  requirements in these two channels.
The main source of the peaking background in the \jpsi channel is the~partially 
reconstructed ${\Bp\to\jpsi\Kstarp}$ decays followed by ${\Kstarp\to\Kp\piz}$ where one photon from 
the \piz decay is not detected. 
In the~\psipr channel the peaking background arises from partially 
reconstructed ${\B\to\psipr\Kp\PY}$ decays combined with a random photon, 
where \B denotes a~\bquark~hadron and $\PY$ denotes additional particles of the \B decay. 
These background contributions are modelled in the fit using non-parametric kernel probability density 
functions~\cite{Keys}, obtained from simulation
of \B~decays to final states containing a~\jpsi or \psipr~meson.  


Combinatorial background is modelled as the product of an~exponential 
function of the $\Ppsi\g\Kp$ invariant mass and a second-order polynomial 
function of the $\jpsi\g$ invariant mass or a third-order polynomial function 
of the $\psipr\g$ invariant mass. For the~latter case, 
the polynomial function is constrained to account 
for the~small available phase space, allowing only two polynomial 
degrees of freedom to vary in the fit.

\begin{figure}[t]
  \setlength{\unitlength}{1mm}
  \centering
  \begin{picture}(140,125) 
    \put(0,65){
      \includegraphics*[width=140mm,
      ]{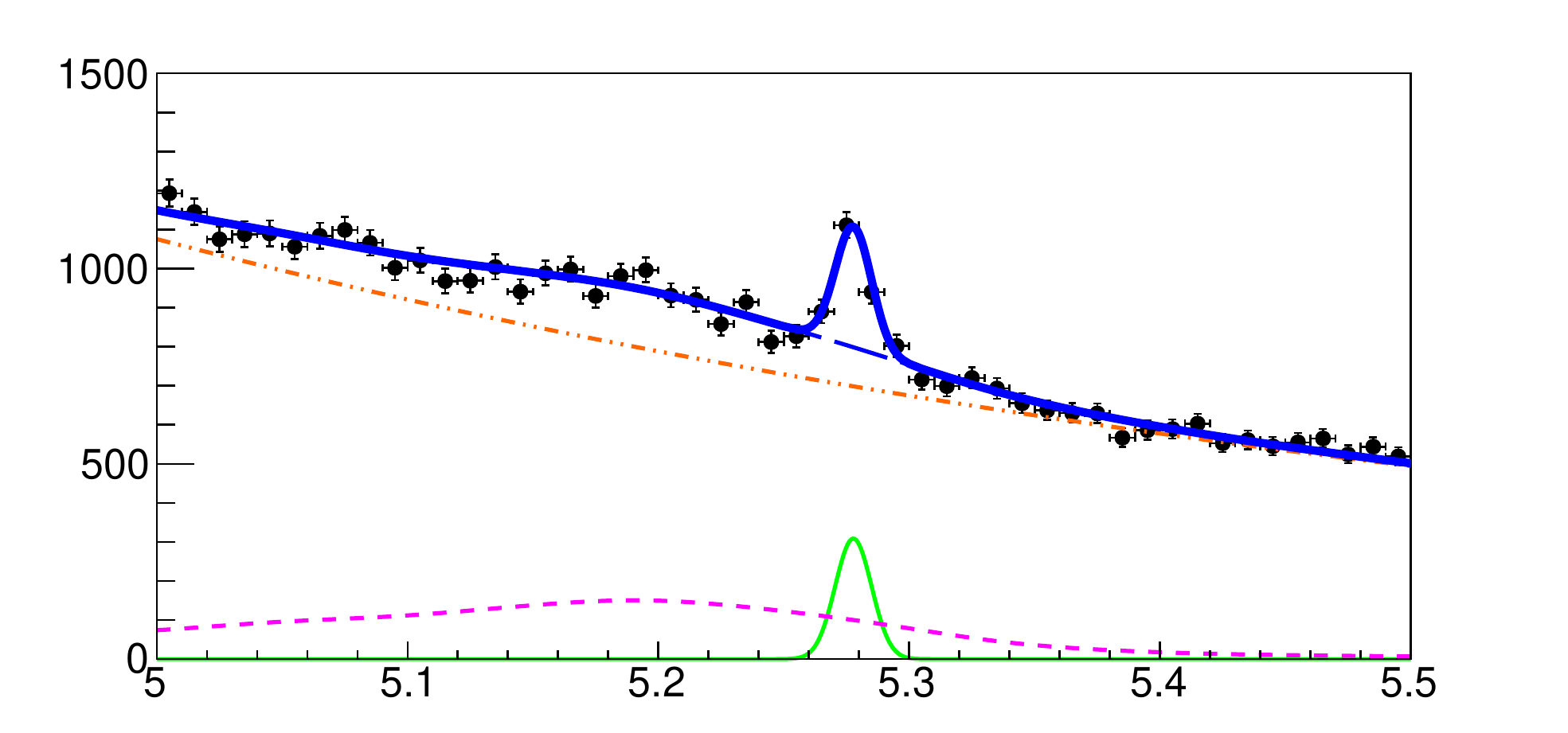}
    } 
    \put(0,0){
      \includegraphics*[width=140mm,
      ]{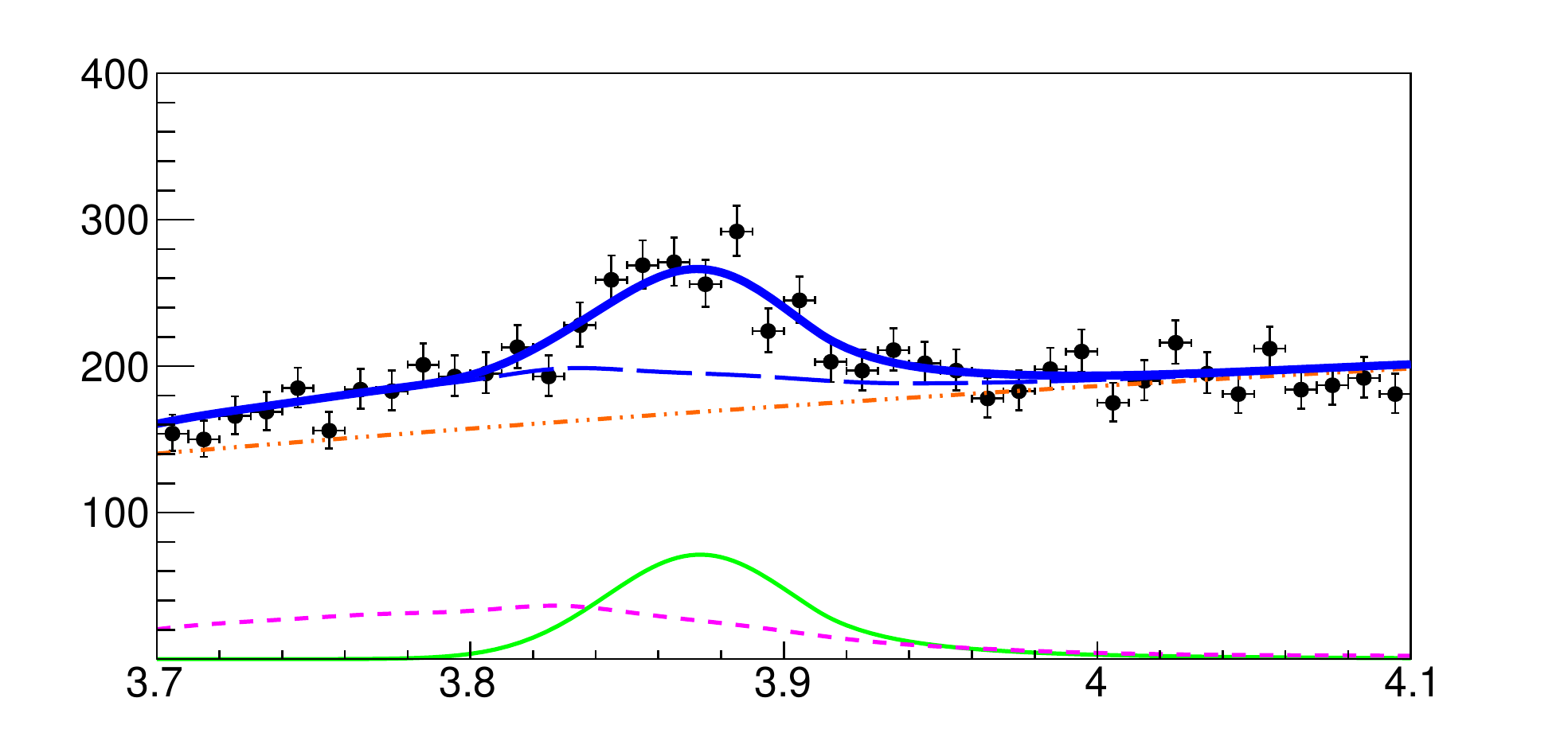}
    }
    \put(-2, 15 ) { \small
      \begin{sideways}%
        Candidates/(10\mevcc) 
      \end{sideways}%
    }
    \put(-2, 80 ) { \small
      \begin{sideways}%
        Candidates/(10\mevcc)
      \end{sideways}%
    } 
    \put(60,64)   { \large  $m_{\jpsi\g\Kp}$ }
    \put(110,64)   { $\left[ \mathrm{GeV}/c^2\right]$}
    \put(60,-1)  { \large  $m_{\jpsi\g}$}
    \put(110,-1)  { $\left[ \mathrm{GeV}/c^2\right]$}
    \put( 112, 112  ){ LHCb}
    \put( 112, 47  ){ LHCb}
    \put( 119, 117  ){ a)}
    \put( 119, 52  ){ b)}
  \end{picture}
  \caption { \small a) Distribution of the ${\jpsi\g\Kp}$ invariant mass with fit projection overlaid, restricted to those candidates with ${\jpsi\g}$ invariant mass
    within $\pm 3\sigma$ from the \X peak position. b)~Distribution of the ${\jpsi\g}$ invariant mass with fit projection overlaid, restricted to those candidates
    with ${\jpsi\g\Kp}$ invariant mass within $\pm 3\sigma$ from the \Bp peak position.
    The total fit (thick solid blue) together with the signal (thin solid green) and background components (dash-dotted orange for the combinatorial,
    dashed magenta for the peaking component and long dashed blue for their sum) are shown. }
   \label{fig:fit2d_jpsi}
\end{figure}

\begin{figure}[t]
  \setlength{\unitlength}{1mm} 
  \centering
   \begin{picture}(140,125)
    \put(0,65){
      \includegraphics*[width=140mm,
      ]{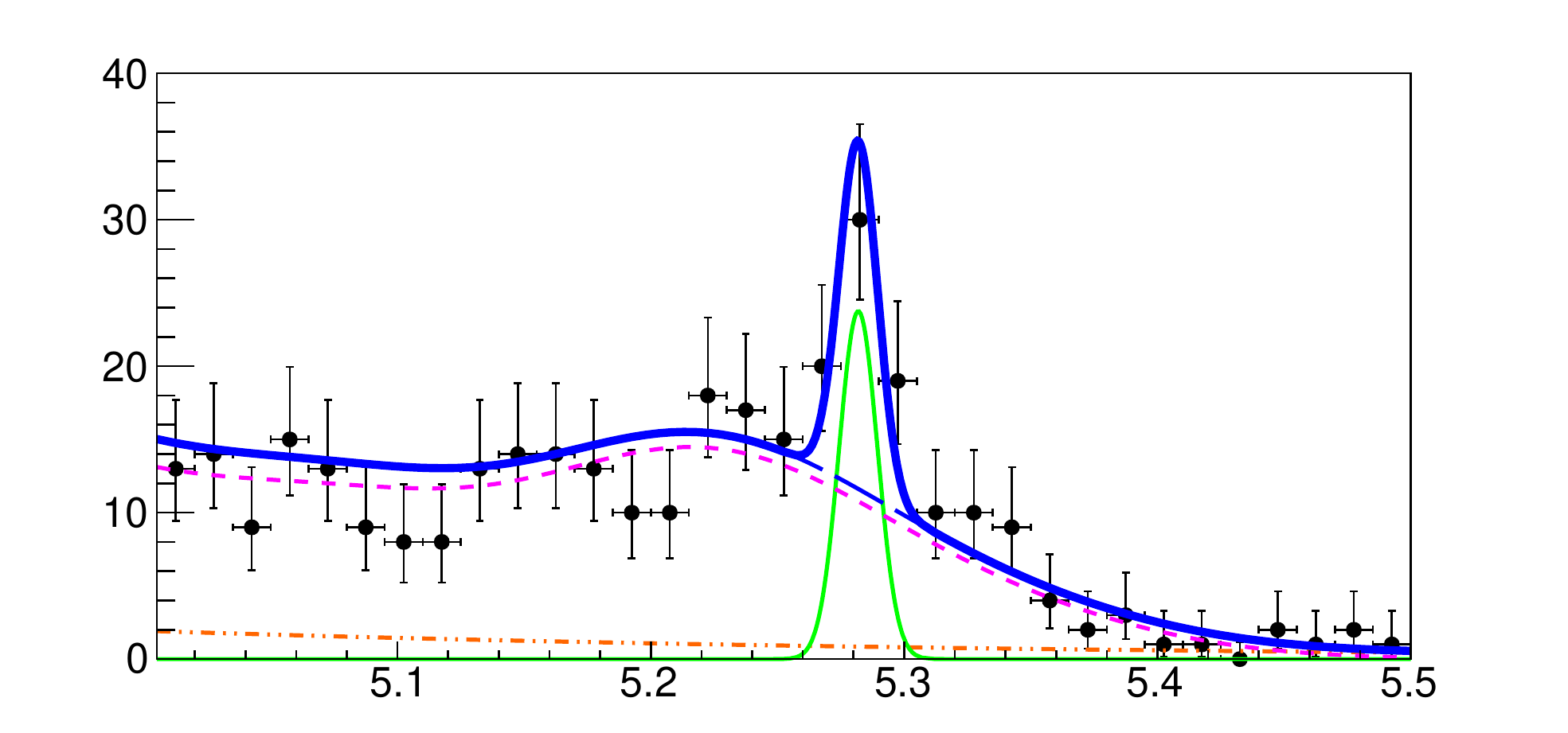}
    }
    \put(0,0){
      \includegraphics*[width=140mm,
      ]{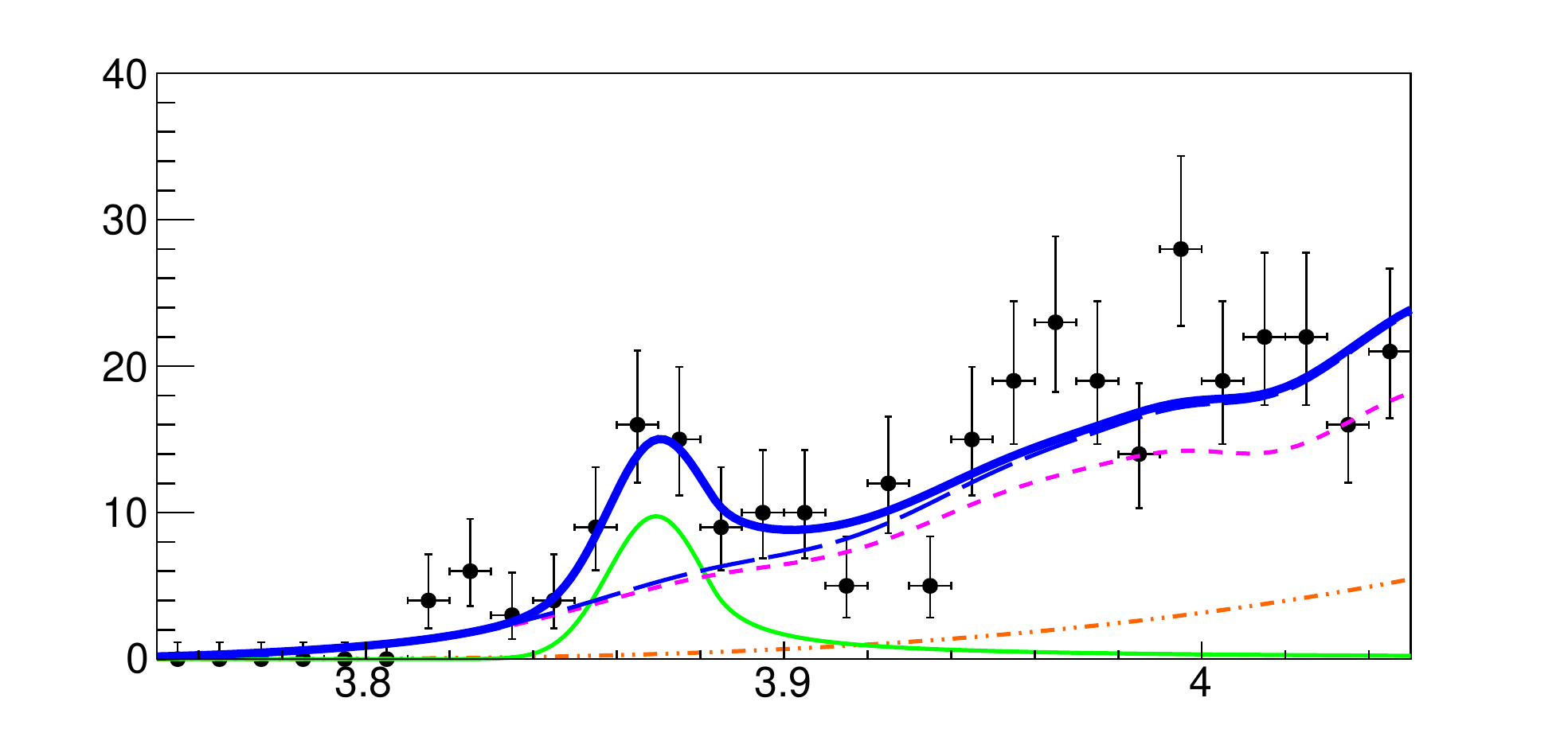}
    }
    \put(-2, 15 ) { \small
      \begin{sideways}%
        Candidates/(10\mevcc)
      \end{sideways}%
    }
    \put(-2, 80 ) { \small
      \begin{sideways}%
        Candidates/(15\mevcc)
      \end{sideways}%
    } 
    \put(60,64)   { \large $m_{\psipr\g\Kp}$ }
    \put(110,64)   { $\left[ \mathrm{GeV}/c^2\right]$}
    \put(60,-1)  { \large $m_{\psipr\g}$}
    \put(110,-1)  { $\left[ \mathrm{GeV}/c^2\right]$}
    \put( 20, 112  ){ LHCb}
    \put( 20, 47  ){ LHCb}
    \put( 20, 117  ){ a)}  
    \put( 20, 52  ){ b)}

    %
    %

  \end{picture}
   \caption { \small a)~Distribution of the ${\psipr\g\Kp}$ invariant mass with fit projection overlaid, restricted to those candidates with ${\psipr\g}$ invariant mass
     within $\pm 3\sigma$ from the \X peak position. 
     b)~Distribution of the ${\psipr\g}$ invariant mass with fit projection overlaid, restricted to those candidates
     with ${\psipr\g\Kp}$ invariant mass within $\pm 3\sigma$ from the \Bp peak position. The~total fit (thick solid blue) together with
     the signal (thin solid green) and background components (dash-dotted orange for the combinatorial, dashed magenta for the peaking component and long dashed blue for their sum) are shown.
   }
  \label{fig:fit2d_psi2s} 
\end{figure}

The fit results for the position of the \Bp~and \X~mass peaks, $m_{\Bp}$ and $m_{\X}$, respectively, 
and the~signal yields $N_{\Ppsi}$ are listed in Table~\ref{table:datafit}.
Projections of the fit on $\Ppsi\g\Kp$ and $\Ppsi\g$ invariant masses are shown in Figs.~\ref{fig:fit2d_jpsi} 
and~\ref{fig:fit2d_psi2s} for the \jpsi and \psipr channels, respectively. 

\begin{table}[t]
\caption{ 
  \small 
  Parameters of the signal functions of the fits to the two-dimensional mass distributions of the ${\Bp\to\X\Kp}$ decays followed by ${\X\to\Ppsi\g}$. 
  Uncertainties are statistical only. 
}
\label{table:datafit}
\centering
\begin{tabular}{lcc}
  
  \multirow{2}{*}{Parameter}	& \multicolumn{2}{c}{Decay mode}	\\
  & $\X\to\jpsi\g$	& $\X\to\psipr\g$                \\
  \hline
  $m_{\Bp}~~~~\,~\left[\!\mevcc\right]$	& $5277.7\pm0.8$			& $5281.9\pm2.4$        \\
  $m_{\X}~\left[\!\mevcc\right]$	& $3873.4\pm3.4$			& $3869.5\pm3.4$        \\
  $N_{\Ppsi}$                           	& $\phantom{0.0}591\pm48\phantom{.}$	& $\phantom{00}36.4\pm9.0$    \\  


\end{tabular}
\end{table}


The significance of the observed signal in the \psipr channel is determined by simulating a~large number of background-only experiments, 
taking into account all~uncertainties in the shape of the background distribution.
The probability for the background to fluctuate to at least the number of observed events is found to be $1.2\times10^{-5}$, corresponding to a 
significance of 4.4 standard deviations for the ${\Bp\to\X\Kp}$ decay followed by ${\X\to\psipr\g}$.



\section{Efficiencies and systematic uncertainties}
\label{sec:Measurement}

The ratio of the $\X\to\psipr\g$ and $\X\to\jpsi\g$ branching fractions is calculated using the formula
\begin{equation}
  R_{\Ppsi\g} = \dfrac{N_{\psipr}}{N_{\jpsi}} \times \dfrac{\Pvarepsilon_{\jpsi}}{\Pvarepsilon_{\psipr}}
  \times \dfrac{\BR(\jpsi\to\mumu)}{\BR(\psipr\to\mumu)} ,
\label{eq:RatioBr}
\end{equation}
where $N_{\jpsi}$ and $N_{\psipr}$ are the measured yields listed in Table~\ref{table:datafit},
and $\Pvarepsilon_{\jpsi}$ and $\Pvarepsilon_{\psipr}$ are the~total efficiencies.
For the ratio of the $\Ppsi\to\mumu$ branching fractions, lepton universality is assumed and a ratio 
of dielectron branching fractions
equal to $7.60 \pm 0.18$~\cite{PDG2012} is used. The~uncertainty is treated as a~systematic uncertainty.

The total efficiency is the product of the geometrical acceptance, the detection, reconstruction,
selection and trigger efficiencies. 
The efficiencies are estimated using simulated events that have been corrected to reproduce
the observed kinematics of \Bp mesons using the high-yield decay 
${\Bp\to\chicone\Kp}$ with ${\chicone\to\jpsi\g}$, which has a topology and kinematics similar
to those of the decays under study.
The ratio of the efficiencies is found to be 
${\Pvarepsilon_{\jpsi}}/{\Pvarepsilon_{\psipr}} = 5.25 \pm 0.04$, where the uncertainty is due to finite size of the simulated samples. 
The ratio of efficiencies is different from unity mainly because of the different photon spectra
in the decays with \jpsi and \psipr in the final state.

Most sources of systematic uncertainty cancel in the ratio, in particular those related to the kaon, muon and $\Ppsi$ reconstruction and 
identification. The remaining systematic uncertainties are summarized in Table~\ref{table:SysUnc} and discussed in turn in the following.

\begin{table}[t]
\caption{ \small Relative systematic uncertainties on the ratio of branching fractions ($R_{\Ppsi\g}$).}
\label{table:SysUnc}
\centering
\begin{tabular}{lc}
	Source 		& Uncertainty~[\%]	\\
	\hline
	$\X\to\jpsi\gamma$ yield determination    & $6$	\\ %
	$\X\to\psitwos\gamma$ yield determination & $7$	\\ %
	Photon reconstruction			& $6$	\\ %
	\Bp kinematics				& $3$	\\ %
	Selection criteria			       & $2$	\\ 
	Trigger					       & $1$	\\ %
	$\BR(\jpsi\to\epem)/\BR(\psipr\to\epem)$ & $2$	\\ %
        Simulation sample size                   & $1$  \\   
	\hline
	Sum in quadrature			& $12$	\\ %
\end{tabular}
\end{table}


Systematic uncertainties related to the signal yield determination are considered in four categories:
signal, 
peaking background, combinatorial background and intervals used in the~fit.
For each category individual uncertainties are estimated using a number of alternative fit models. 
The maximum deviations from the baseline values of the yields are taken as individual systematic uncertainties, 
which are then added in quadrature.
The systematic uncertainties on the event yields are dominated by uncertainties in the description of backgrounds 
and are 6\,\% and 7\,\% in the \jpsi and \psipr channels, respectively. 



Another important source of systematic uncertainty arises from the potential disagreement between  
data and simulation in the estimation of efficiencies. 
This includes the photon reconstruction efficiency, 
the imperfect knowledge of \Bp kinematics and the~description of the~selection criteria efficiencies. 
The photon reconstruction efficiency is studied using a large sample of $\Bp\to\jpsi\Kstarp$ decays, followed by
$\Kstarp\to\Kp\piz$ and $\piz\to\g\g$ decays.
The relative yields of  $\Bp\to\jpsi\Kstarp$ and  $\Bp\to\jpsi\Kp$
decays are compared in data and simulation.
For photons with transverse momentum greater than 0.6\gevc, the~agreement between data and
simulation is within 6\,\%, which is assigned as the systematic uncertainty due
to the photon reconstruction.

The systematic uncertainty related to the knowledge of the \Bp production properties is estimated by comparing the ratio of efficiencies determined without 
making corrections to the \Bp transverse momentum and rapidity spectra to the default ratio of efficiencies determined after the corrections. The relative difference
between the two methods is found to be~3\,\% and is conservatively assigned as the systematic uncertainty from this source. 


To study the uncertainty due to selection criteria, the high-yield  
decay ${\Bp\to\chicone\Kp}$, followed by ${\chicone\to\jpsi\g}$, which has a similar topology to the decays 
studied in this analysis, is used.
The selection criteria for the photon and kaon transverse momentum, the $\piz\to\g\g$ veto and the \chisq/ndf of the kinematic fit are studied.
The selection criteria are varied in ranges corresponding to as much as a $30\,\%$ change in the signal yields and the ratios of the selection and 
reconstruction efficiencies are compared between data and simulation. 
The largest difference of 2\,\% is assigned as the corresponding systematic uncertainty. 


The systematic uncertainty related to the trigger efficiency is obtained by comparing the~trigger 
efficiency ratios in data and simulation for the~high yield decay modes $\Bp\to\jpsi\Kp$ and 
$\Bp\to\psitwos\Kp$, which have similar kinematics and the same trigger requirements as the channels 
under study in this analysis~\cite{LHCb-PAPER-2012-010}. 
An~agreement within 1\,\% is found, which is assigned as the corresponding systematic uncertainty. 







\section{ Results and summary}
\label{sec:Result}

Using a sample of $\proton\proton$ collisions at centre-of-mass energies of 7 and 8\tev, corresponding to an integrated luminosity of 3\invfb, evidence 
for the decay $\X\to\psipr\g$ in \mbox{$\Bp\to\X\Kp$} decays is found with a significance of 4.4 standard deviations.
Its~branching fraction, normalized to that of the $\X\to\jpsi\g$ decay mode is measured \mbox{to be} 
\begin{equation*}
R_{\Ppsi\g} = \dfrac{ \BR(\X\to\psipr\g)}{ \BR(\X\to\jpsi\g )} = 2.46\pm0.64\pm0.29, 
\end{equation*}
\noindent where the first uncertainty is statistical and the second is systematic.
This result is compatible with, but more precise than, previous measurements~\cite{babar,belle}. 
The measured value of $R_{\Ppsi\g}$ agrees with expectations for a pure charmonium interpretation
of the \X state~\cite{BarnesGodfreySwanson,BarnesGodfrey,LiChao,Lahde,Simonov,identifyX1,identifyX2}
and a~molecular-charmonium mixture interpretations~\cite{Simonov,EichtenLaneQuigg}.
However, it does not support a pure~$\D\Dstarb$~molecular interpretation~\cite{Swanson,DongFaeser,FerrettiGalata} of the \X~state.

\section*{Acknowledgements}

\noindent We express our gratitude to our colleagues in the CERN
accelerator departments for the excellent performance of the LHC. We
thank the technical and administrative staff at the LHCb
institutes. We acknowledge support from CERN and from the national
agencies: CAPES, CNPq, FAPERJ and FINEP (Brazil); NSFC (China);
CNRS/IN2P3 and Region Auvergne (France); BMBF, DFG, HGF and MPG
(Germany); SFI (Ireland); INFN (Italy); FOM and NWO (The Netherlands);
SCSR (Poland); MEN/IFA (Romania); MinES, Rosatom, RFBR and NRC
``Kurchatov Institute'' (Russia); MinECo, XuntaGal and GENCAT (Spain);
SNSF and SER (Switzerland); NASU (Ukraine); STFC and the Royal Society (United
Kingdom); NSF (USA). We also acknowledge the support received from EPLANET, 
Marie Curie Actions and the ERC under FP7. 
The Tier1 computing centres are supported by IN2P3 (France), KIT and BMBF (Germany),
INFN (Italy), NWO and SURF (The Netherlands), PIC (Spain), GridPP (United Kingdom).
We are indebted to the communities behind the multiple open source software packages on which we depend.
We are also thankful for the computing resources and the access to software R\&D tools provided by Yandex LLC (Russia).

\addcontentsline{toc}{section}{References}
\bibliographystyle{LHCb}
\bibliography{main,LHCb-PAPER,LHCb-CONF,LHCb-DP,local}

\end{document}